\documentclass[12pt]{article}

\usepackage{amsfonts}
\usepackage{amsmath}
\usepackage{amscd}
\usepackage{amssymb}
\usepackage{amsthm}
\usepackage{amsxtra}

\newtheorem{prop}{Proposition}

\theoremstyle{definition} 
\newtheorem{rem}{Remark} 
\newtheorem{exa}{Example}

\begin{document}

\title{Dynamics of wave packets in the functional mechanics}
\author
          {Zharinov V.V.
             \thanks{Steklov Mathematical Institute}
             \thanks{E-mail: zharinov@mi.ras.ru}}
\date{}
\maketitle

\begin{abstract}
In these notes we consider the mathematical aspects 
of the functional mechanics proposed and developed by 
I.~V. Volovich. 
\end{abstract} 
{\bf keywords:} functional mechanics, wave packet, Liouville equation, 
dispersion, Hamiltonian dynamics, Lagrangian dynamics, divergence free dynamics.

\section{Preliminaries.} 

In these notes we consider mathematical aspects 
of the functional mechanics proposed and developed  
by I.~V. Volovich in his papers \cite{IVV1}--\cite{IVV5}.
We treat only formal analytical problems, in particular, 
we assume that all functions are sufficiently 
smooth and all integrals are convergent. 
The relationship of the functional mechanics with fuzzy logic 
was discussed in the article \cite{Z}. 

We use the notation:
\begin{itemize} 
	\item 
		the summation over repeated upper and lower indices 
		in the prescribed limits is assumed. 
	\item 
		$\mathbb N=\{1,2,\dots\}, \ \mathbb Z_+=\{0,1,2,\dots\}, \ 
		\mathbb Z=\{0,\pm1,\pm2,\dots\}$; 
	\item 
		$\mathbb R_+=\{x\in\mathbb R\mid x\ge0\}$; 
	\item 
		$\langle a,b\rangle=\sum_\mu a^\mu b^\mu$ for all 
		$a=(a^\mu),b=(b^\mu)\in\mathbb R^m$, $\mu\in\overline{1,m}$, 
		$m\in\mathbb N$, in particular, 
		$a^2=\langle a,a\rangle=\sum_\mu(a^\mu)^2$; 
	\item 
		$D_t=\partial_t+v^\mu(x)\partial_{x^\mu}$ 
		is the total time derivative along a vector field 
		$v=(v^\mu) : \mathbb R^m\to\mathbb R^m$, 
		$x=(x^\mu)\mapsto v(x)=(v^\mu(x))$; 
	\item 
		$\dot\phi(t)=\frac d{dt}\phi(t)$ for any function 
		$\phi(t)$.  
\end{itemize} 

\section{Wave packets.} 
A wave packet is a mapping 
\begin{equation}
	R : \mathbb R\times\mathbb R^m\to\mathbb R_+, \quad (t,x)\mapsto R(t,x), 
	\quad m\in\mathbb N. 
\end{equation}
In the functional mechanics it is assumed that the wave packet 
satisfies the Liouville equation 
\begin{equation}\label{LE} 
	\partial_t R+\partial_{x^\mu}\big(v^\mu\cdot R\big)=
	D_tR+\mathrm{div\,}v\cdot R=0,
\end{equation}
where $v : \mathbb R^m\to\mathbb R^m$ is a given vector field, 
and $\mathrm{div}\,v=\partial_{x^\mu}v^\mu(x)$ is its divergence. 

The Liouville equation (\ref{LE}) describes the conservation law 
for the pair $(R,vR)$, hence, the integral 
$\int R(t,x)d^m x=\mathrm{const}$, 
and one may assume that it is equal to 1. 
Indeed, 
\begin{equation*} 
	\frac d{dt}\int R(t,x)d^mx=\int\partial_tR(t,x)d^mx
	=-\int\partial_{x^\mu}\big(v^\mu(x)R(t,x)\big)d^mx=0,
\end{equation*} 
assuming $|v(x)R(t,x)|\to 0$ sufficiently fast, 
when $|x|\to\infty$. 

Below, we assume $\int R(t,x)d^mx=1$ for all $t\in\mathbb R$. 
In this case, the average value of a function $f : \mathbb R\times\mathbb R^m\to\mathbb R$, 
$(t,x)\mapsto f(t,x)$, is defined as 
\begin{equation}\label{AV} 
	\overline f(t)=\int f(t,x)R(t,x)d^mx, \quad t\in\mathbb R. 
\end{equation} 
We assume, also, that the wave packet $R(t,x)$ satisfies 
the Liouville equation (\ref{LE}). Then the time derivative 
\begin{align*}  
	\dot{\overline f}(t)
	&=\int\big(\partial_tf(t,x)\cdot R(t,x)
	+f(t,x)\cdot\partial_tR(t,x)\big)d^mx=
	(\text{see }(\ref{LE}))  \\
	&=\int\big(\partial_t f\cdot R-f\cdot\partial_{x^\mu}(v^\mu\cdot R)\big)
	d^mx=(\text{integration by parts})  \\ 
	&=\int\big(\partial_tf+v^\mu\partial_{x^\mu}f\big)R \, d^mx
	=\int D_tf\cdot R\, d^mx.
\end{align*} 
We summarize these considerations as the following 

\begin{prop} 
If a wave packet $R(t,s)$ satisfies the Liouville equation 
(\ref{LE}), then the time derivative  
\begin{equation*} 
	\dot{\overline f}(t)=\int D_tf\cdot R\,d^mx, \quad t\in\mathbb R.
\end{equation*}
In particular, $\overline f(t)=\mathrm{const}$, 
if $D_t f(t,x)=0$. 
\end{prop}

We are most interested in the average values of the coordinates  
\begin{equation*} 
	\overline{x^\mu}(t)=\int x^\mu R(t,x)d^mx, 
	\quad \mu\in\overline{1,m},
\end{equation*}
and the dispersion 
\begin{equation*} 
	\overline{\Delta x^2}(t)=\int(\Delta x)^2R(t,x)d^mx
	=\overline{x^2}(t)-\big(\overline x(t)\big)^2, 
\end{equation*} 
where 
$\Delta x=x-\overline x(t)$, \  
$\overline{x^2}(t)=\int x^2R(t,x)d^mx$. 

\begin{prop} 
In the above settings, 
\begin{itemize} 
	\item 
		$\dot{\overline x}(t)=\overline v (t)$, \ where \ 
		$\overline v(t)=\int v(x)R(t,x)d^mx$; 
	\item 
		$\dot{\overline{\Delta x^2}}(t)
		=2\overline{\langle\Delta x,\Delta v\rangle}(t)
		=2\big(\overline{\langle x,v\rangle}(t)
		-\langle\overline x(t),\overline v(t)\rangle\big)$, 
		$\Delta v=v-\overline v(t)$.
\end{itemize}
\end{prop}

\section{Hamiltonian dynamics.} 
In the Hamiltonian situation, $m=2n$, 
$x=(q,p)\in\mathbb R^n\times\mathbb R_n$, the dynamics is governed 
by a Hamiltonian $H=H(q,p)$, the vector field 
\begin{equation*}  
	v=v(q,p)=\big(V,F\big) : \mathbb R^n\times\mathbb R_n
	\to\mathbb R^n\times\mathbb R_n,
\end{equation*}
where 
$V=\big(V^\alpha=\partial_{p_\alpha}H\big)$ is the generalized velocity, 
$F=\big(F_\alpha=-\partial_{q^\alpha}H\big)$ is the generalized force,
$\alpha\in\overline{1,n}$, the total time derivative 
\begin{equation*}
	D_t=\partial_t+V^\alpha\partial_{q^\alpha}+F_\alpha\partial_{p_\alpha}. 
\end{equation*}
\begin{prop}\label{DF} 
In the Hamiltonian dynamics, the divergence 
\begin{equation*} 
	\partial_{x^\mu}v^\mu=\partial_{q^\alpha}(\partial_{p_\alpha}H)+
	\partial_{p_\alpha}(-\partial_{q^\alpha}H)=0, 
\end{equation*}  
and the Liouville equation (\ref{LE}) has the familiar look 
\begin{equation}\label{HE} 
	\partial_tR+\big\{H,R\big\}=0, \qquad 
	\big\{H,R\big\}=\partial_{p_\alpha}H\cdot\partial_{q^\alpha}R
	-\partial_{q^\alpha}H\cdot\partial_{p_\alpha}R.
\end{equation} 
\end{prop}

In particular, if $H=\frac{p^2}{2m}+U(q)$, 
$p^2=\langle p,p\rangle=\sum_\alpha(p_\alpha)^2$, 
then 
\begin{equation*} 
	\partial_tR+\{H,R\}=\partial_t+\sum_\alpha\bigg(
	\frac 1m p_\alpha\cdot\partial_{q^\alpha}R
	-\partial_{q^\alpha}U\cdot\partial_{p_\alpha}R\bigg).
\end{equation*}

Here (we set $dq=d^nq$, $dp=d^np$, for brevity), 
\begin{align*}  
	&\overline q(t)=\int qR(t,q,p)dqdp, \quad  
	\dot{\overline q}(t)=\int D_tq\cdot R(t,q,p)dqdp=\overline V(t); \\  
	&\overline p(t)=\int pR(t,q,p)dqdp, \quad  
	\dot{\overline p}(t)=\int D_tp\cdot R(t,q,p)dqdp=\overline F(t). 
\end{align*}
Thus, in the functional Hamiltonian dynamics we have 
the familiar equations for the average values 
\begin{equation*}
	\dot{\overline q}=\overline{\partial_pH}=\overline V, 
	\qquad 
	\dot{\overline p}=-\overline{\partial_qH}=\overline F. 
\end{equation*} 
Further, 
\begin{align*} 
		\overline{\Delta q^2}(t)
	&=\int\Delta q^2R(t,q,p)dqdp
		=\int(q-\overline q(t))^2R(t,q,p)dqdp; \\
		\dot{\overline{\Delta q^2}}(t)
	&=\int D_t(q-\overline q(t))^2\cdot R(t,q,p)dqdp \\
	&=2\int\langle(q-\overline q(t)),D_t(q-\overline q(t))\rangle 	
		\cdot R(t,q,p)dqdp \\ 
	&=2\int\langle q-\overline q(t),V-\overline V(t)\rangle 
		\cdot R(t,q,p)dqdp.	
\end{align*} 
Thus, 
\begin{equation*} 
	\dot{\overline{\Delta q^2}}(t)=
	2\overline{\langle\Delta q,\Delta V\rangle}(t), \quad 
	\Delta q=q-\overline q(t), \ \Delta V=V(q)-\overline V(t).
\end{equation*} 
In the same way, 
\begin{align*} 
		\overline{\Delta p^2}(t)
	&=\int\Delta p^2R(t,q,p)dqdp
		=\int(p-\overline p(t))^2R(t,q,p)dqdp; \\
		\dot{\overline{\Delta p^2}}(t)
	&=2\overline{\langle\Delta p,\Delta F\rangle}(t), \quad 
		\Delta p=p-\overline p(t), \ \Delta F=F(q)-\overline F(t).
\end{align*} 
\begin{prop} 
If the Hamiltonian $H(q,p)=T(p)+U(q)$, 
where $T(p)$ is the kinetic energy, $U(q)$ is the potential energy, 
while the wave packet is factorised, i.e., 
\begin{equation*}
	R(t,q,p)=Q(t,q)\cdot P(t,p), \quad 
	\int Q(t,q)dq=\int P(t,p)dp=1. 
\end{equation*} 
Then, 
\begin{equation*} 
	\dot{\overline{\Delta q^2}}(t)=\dot{\overline{\Delta p^2}}(t)=0, 
	\quad t\in\mathbb R.
\end{equation*} 
\end{prop}
Indeed, in this case, 
\begin{align*} 
	&\overline{\langle\Delta q,\Delta V\rangle}(t)
		=\bigg\langle\int\Delta q\cdot Q(t,q)dq,
		\int\Delta V\cdot P(t,p)dp\bigg\rangle=
		\langle\overline{\Delta q},\overline{\Delta V}\rangle=0, \\
	&\overline{\langle\Delta p,\Delta F\rangle}(t)
		=\bigg\langle\int\Delta p\cdot P(t,p)dp,
		\int\Delta F\cdot Q(t,q)dq\bigg\rangle
		=\langle\overline{\Delta p},\overline{\Delta F}\rangle=0,
\end{align*}
because here, for example, 
\begin{align*} 
	&\overline q(t)=\int qQ(t,q)dq\cdot\int P(t,p)dp
		=\int qQ(t,q)dq, \\
	&\overline{\Delta q}(t)
		=\int(q-\overline q(t))Q(t,q)dq
		=\overline q(t)-\overline q(t)=0.
\end{align*} 

\begin{rem} 
The problem here is the existence of a suitable 
factorised wave packet satisfying the Liouville equation (\ref{LE}).
\end{rem} 

\section{Lagrangian dynamics.}\label{LD}
In the Lagrangian dynamics, $m=2n$, $x=(q,\dot q)\in\mathbb R^n\times\mathbb R^n$, 
the dynamics is governed by a Lagrangian $L=L(q,\dot q)$. 
The total time derivative 
\begin{equation*} 
	D_t=\partial_t+\dot q^\alpha\partial_{q^\alpha}+\ddot q^\alpha\partial_{\dot q^\alpha},
\end{equation*} 
where the coefficients 
$\ddot q^\alpha(q,\dot q)$, $\alpha\in\overline{1,n}$,  
should be found from the variational equations (Euler-Lagrange equations) 
\begin{equation*} 
	\delta_{q^\alpha}L=\partial_{q^\alpha}L-\big(\partial_{\dot q^\alpha}L\big)^.
	=\partial_{q^\alpha}L-\partial^2_{q^\beta\dot q^\alpha}L\cdot
	\dot q^\beta-\partial^2_{\dot q^\beta\dot q^\alpha}L\cdot
	\ddot q^\beta=0.
\end{equation*} 
The crucial role here is played by the Hessian 
$H=\det\big(\partial^2_{\dot q^\alpha\dot q^\beta}L\big)$. 
If the Hessian is non-degenerate then the function $\ddot q$ is unequivocally defined by the variational equations, 
in the opposite case one should impose some additional constraints.  
Note, if the Hessian is not-degenerate one can pass from the variables $q,\dot q$ to the variables $q,p$, from the Lagrangian $L(q,\dot q)$ 
pass to the Hamiltonian $H(q,p)$, and come to the Hamiltonian dynamics. 

After the above problem is solved, the vector field takes the form 
$v=(\dot q,\ddot q(q,\dot q))$, and the Liouville equation for the 
wave packet $R(t,q,\dot q)$ is written as  
\begin{equation*} 
	\partial_tR+\partial_{q^\alpha}\big(\dot q^\alpha\cdot R\big)
	+\partial_{\dot q^\alpha}\big(\ddot q^\alpha\cdot R\big)=
	D_t R+\mathrm{div}\,v\cdot R=0, 
\end{equation*}
where the divergence $\mathrm{div}\,v=\partial_{q^\alpha}\dot q^\alpha
+\partial_{\dot q^\alpha}\ddot q^\alpha=0+\partial_{\dot q^\alpha}\ddot q^\alpha
=\partial_{\dot q^\alpha}\ddot q^\alpha$.
 
\section{Divergence free dynamics.}\label{DFT} 
Here we suppose that the vector field $v : \mathbb R^m\to\mathbb R^m$ 
is divergence free, i.e., $\mathrm{div\,}v=0$. 
In this case, the Liouville equation takes the form 
\begin{equation}\label{LDF} 
	D_tR=\partial_tR+v^\mu\partial_{x^\mu}R=0,
\end{equation} 
and according to the theory of the equations in the 
partial derivatives of the first order, has the general solution 
\begin{equation*}
	R(t,x)=\rho(S(t,x)), \quad R(0,x)=\rho(x), 
\end{equation*} 
where the profile $\rho : \mathbb R^m\to\mathbb R_+$, $s\mapsto\rho(s)$, 
is arbitrary, while the flow 
\begin{equation*}
	S : \mathbb R\times\mathbb R^m\to\mathbb R^m, \quad (t,x)\mapsto s=S(t,x), 
	\quad S(0,x)=x. 
\end{equation*} 
In more detail, $S=(S^\mu(t,x))$, where the components $S^\mu(t,x)$, 
$\mu\in\overline{1,m}$, are the first integrals of the equation (\ref{LDF}).

We also assume that $\int\rho(S(t,x))d^mx=1$ for all $t\in\mathbb R$. 
In this case, the average value (\ref{AV}) reduces to 
\begin{equation*} 
	\overline f(t)=\int f(t,x)\rho(S(t,x))d^mx, \quad t\in\mathbb R. 
\end{equation*} 

If the induced mapping $S(t): \mathbb R^m\to\mathbb R^m$, 
$x\mapsto s=S(t,x)$, is invertible for every $t\in\mathbb R$, i.e., 
there exists the inverse mapping $X(t) : \mathbb R^m\to\mathbb R^m$, 
$s\mapsto x=X(t,s)$, and the Jacobian 
$J(t,s)=\det\|\partial_{s^\mu}X^\nu(t,s)\|\ne0$, 
then the average value can be written as follows 
\begin{equation*} 
	\overline f(t)=\int f(t,X(t,s))J(t,s)\cdot\rho(s)d^ms, 
	\quad t\in\mathbb R. 
\end{equation*} 

\begin{prop}\label{SC} 
In the above settings, let 
\begin{equation*} 
	S=A(t)x+a(t), \quad D_tS=\dot A(t) x+\dot a(t)+A(t)v(x)=0, 
\end{equation*} 
where 
\begin{itemize} 
	\item  
		$s=(s^\mu)\in\mathbb R^m$, \ $S=(S^\mu(t,x)) :
		\mathbb R\times\mathbb R^m\to\mathbb R^m$, 
	\item  
		$A=(A^\mu_\nu(t))\in\mathbb R^m_m$, \ $x=(x^\mu)\in\mathbb R^m$, 
		\ $a=(a^\mu(t))\in\mathbb R^m$, \ $t\in\mathbb R$, 
		$A(0)$ is the identity matrix, $\det A=1$, $a(0)=0$; 
	\item 
		$v=(v^\mu(x)) : \mathbb R^m\to\mathbb R^m$, $x\mapsto v(x)$, 
		\quad $\mathrm{div\,}v=0$.  
\end{itemize} 
Then 
\begin{itemize} 
	\item 
		the induced mapping $S(t) :\mathbb R^m\to\mathbb R^m$, \ 
		$x\mapsto s=S(t,x)$ is invertible for any $t\in\mathbb R$, 
		i.e there is defined the inverse induced mapping 
		$X(t) : \mathbb R^m\to\mathbb R^m$,
		$s\mapsto x=X(t,s)$, $X(t,s)=B(t)s+b(t)$, 
		where $B=(B^\mu_\nu(t))=A^{-1}$, $b=-Ba$, $J(t,s)=\det B=1$; 
	\item 
		$\overline x(t)=B(t)\overline s+b(t)$, \ 
		$\overline s=\int s\rho(s)d^ms$; 
	\item 
		$\Delta x=x-\overline x(t)=B(t)\Delta s$, \ 
		$\Delta s=s-\overline s$; 
	\item 
		$\Delta x^2=\langle\Delta x,\Delta x\rangle
		=\sum_\lambda B^\lambda_\mu(t)B^\lambda_\nu(t)
		\cdot\Delta s^\mu \Delta s^\nu=
		g_{\mu\nu}(t)\cdot\Delta s^\mu \Delta s^\nu$;  
	\item 
		$\overline{\Delta x^2}=g_{\mu\nu}(t)\cdot 
		\overline{\Delta s^\mu \Delta s^\nu}
		=g_{\mu\nu}(t)\cdot\sigma^{\mu\nu}
		=\mathrm{Tr}(g\sigma)\ge0$; 
	\item 
		$g_{\mu\nu}(t)=\sum_\lambda B^\lambda_\mu(t)B^\lambda_\nu(t)$, 
		\quad 
		$\sigma^{\mu\nu}=\overline{\Delta s^\mu \Delta s^\nu}
		=\overline{s^\mu\cdot s^\nu}
		-\overline{s^\mu}\cdot\overline{s^\nu}$; 
	\item 
		the matrices $g=(g_{\mu\nu}(t))$ and 
		$\sigma=(\sigma^{\mu\nu})$ 
		are positive definite, $g(0)$ is the identity matrix. 
\end{itemize} 
The matrix $g(t)$ characterises the dynamics of the wave packet, 
while the matrix $\sigma$ characterises its original profile.
\end{prop}

\section{Examples.} 
\begin{exa}\label{FM} 
Let 	$n=1$, $H=\frac{p^2}{2m}$, $m>0$, then the total time derivative 
$D_t=\partial_t+\frac pm\partial_q$, the Liouville equation takes the form 
\begin{equation*} 
	D_t R=\partial_tR+\frac pm\partial_qR=0,
\end{equation*} 
and has the general solution $R(t,q,p)=\rho(r,s)$, 
where $\rho=\rho(r,s)$, $r,s\in\mathbb R$, is an arbitrary function, and 
$r=q-\frac pmt, \quad s=p$.  
Thus, here  
\begin{equation*} 
	\begin{cases}r=q-\frac tmp. \\ s=p;\end{cases}
	\qquad 
	\begin{cases}q=r+\frac tms, \\ p=s;\end{cases}
	\qquad J(t;q,p)=1.
\end{equation*} 
In particular, the conditions of Proposition \ref{SC} 
are fulfilled with 
\begin{equation*} 
	x=\begin{pmatrix}q\\ p\end{pmatrix}, \quad 
	s=\begin{pmatrix}r\\ s\end{pmatrix}, \quad 
	A=\begin{pmatrix}1 & -\frac tm \\0 & 1\end{pmatrix},\quad 
	B=\begin{pmatrix}1 & \frac tm \\0 & 1\end{pmatrix},\quad 
	a=b=0.
\end{equation*}

The average value of a function $f(t,q,p$ takes the form 
\begin{equation*} 
	\overline f(t)=\int f\big(t,r+\frac tms,s\big)
		\cdot\rho(r,s)drds. 
\end{equation*} 
In particular, 
\begin{itemize} 
	\item 
		$\overline q(t)=\overline r+\frac tm\overline s, 
		\quad \overline p(t)=\overline s$; 
	\item 
		$\overline{\Delta q^2}(t)=\overline{\Delta r^2}
		+2\frac tm\cdot\overline{\Delta r\cdot\Delta s}
		+\big(\frac tm\big)^2\cdot\overline{\Delta s^2}, 
		\quad \overline{\Delta p^2}(t)=\overline{\Delta s^2}$;  
	\item 
		$\overline r=\int r\rho(r,s)drds$, \quad 
		$\overline s=\int s\rho(rs)drds$; 
	\item 
		$\overline{\Delta r^2}=\int\Delta r^2\rho(r,s)drds$; 
	    \quad 
		$\overline{\Delta s^2}=\int \Delta s^2\rho(r,s)drds$, \\ 
		$\overline{\Delta r\cdot\Delta s}=\int 
		\Delta r\Delta s\,\rho(r,s)drds$, .
\end{itemize} 
So, 
$\overline{\Delta x^2}=\overline{\Delta q^2}+\overline{\Delta p^2}
=g_{\mu\nu}\cdot\sigma^{\mu\nu}=\mathrm{Tr}(g\sigma)$, 
where 
\begin{equation*} 
	g=\begin{pmatrix}
		      1 & \frac tm   \\ 
		\frac tm &  1+\big(\frac tm\big)^2 
	  \end{pmatrix}, \quad 
	 \sigma=\begin{pmatrix}
	        \overline{\Delta r^2}&\overline{\Delta r\cdot\Delta s} \\
	        \overline{\Delta r\cdot\Delta s}&\overline{\Delta s^2}
	        \end{pmatrix}, \quad 
	        \det g=1, \quad \det\sigma>0.
\end{equation*}
\end{exa} 

\begin{exa}\label{AM} 
Let 	$n=1$, $H=\frac{p^2}{2m}-Fq$, $F=const$, 
then the total time derivative $D_t=\partial_t+\frac pm\partial_q+F\partial_p$, 
while the Liouville equation takes the form 
\begin{equation*} 
	D_t R=\partial_tR+\frac pm\partial_qR+F\partial_pR=0,
\end{equation*} 
and has the general solution $R(t,q,p)=\rho(r,s)$, 
where $\rho=\rho(r,s)$, $r,s\in\mathbb R$, is an arbitrary function, and
\begin{equation*} 
	r=q-\frac{pt}m+\frac{Ft^2}{2m}, \quad s=p-Ft. 
\end{equation*} 
Thus, here 
\begin{equation*} 
	\begin{cases}
	r=q-\frac tmp+\frac{Ft^2}{2m}, \\ s=p-Ft; 
	\end{cases}
	\quad
	\begin{cases}
	q=r+\frac tms+\frac{Ft^2}{2m}, \\ p=s+Ft; 
	\end{cases}
	\quad J(t;r,s/q,p)=1.
\end{equation*} 
The average value of a function $f(t,q,p$ takes the form 
\begin{equation*} 
	\overline f(t)
	=\int f\big(t,r+\frac tms+\frac{Ft^2}{2m},s+Ft\big)
		\cdot\rho(r,s)drds. 
\end{equation*} 
In particular, 
\begin{itemize} 
	\item 
		$\overline q(t)
		=\overline r+\frac{\overline st}m+\frac{Ft^2}{2m}$, 
		\quad $\overline p(t)=\overline s+Ft$; 
	\item 
		$\overline{\Delta q^2}(t)=\overline{\Delta r^2}
		+2\overline{\Delta r\cdot\Delta s}\cdot\frac tm
		+\overline{\Delta s^2}\big(\frac tm\big)^2, 
		\quad \overline{\Delta p^2}(t)=\overline{\Delta s^2}$, \\
		like in the previous example, because here $\overline F=F$, 
		so $\Delta F=0$.
\end{itemize}  
\end{exa}

\begin{exa}\label{O} 
Let 	$n=1$, $H=\frac{p^2}{2m}+\frac{kq^2}2$, $k>0$,  
then the total time derivative $D_t=\partial_t+\frac pm\partial_q-kq\partial_p$, 
the Liouville equation takes the form 
\begin{equation*} 
	D_tR=\partial_tR+\frac pm\partial_qR-kq\partial_pR=0,
\end{equation*}
and has the general solution $R(t,q,p)=\rho(r,s)$, 
where $\rho=\rho(r,s)$, $r,s\in\mathbb R$, is an arbitrary function, and
\begin{equation*} 
	r=q\cos\omega t-\frac p{m\omega}\sin\omega t, \quad  
	s=m\omega q\sin\omega t+p\cos\omega t,
\end{equation*}
$\omega=\sqrt{\frac km}$, $m\omega=\sqrt{mk}$. 
Thus, here 
\begin{equation*}
	\begin{cases} 
	r=q\cos\omega t-p\frac1{m\omega}\sin\omega t, \\
	s=q\,m\omega\sin\omega t+p\cos\omega t; 
	\end{cases}
	\quad 
	\begin{cases} 
	q=r\cos\omega t+s\frac1{m\omega}\sin\omega t, \\
	p=-r\,m\omega\sin\omega t+s\cos\omega t. 
	\end{cases}
\end{equation*}
The average value of a function $f(t,q,p$ is written in the form 
\begin{equation*} 
	\overline f(t)
	=\int f\big(t,r\cos\omega t+s\,\frac1{m\omega}\sin\omega t, 
	-r\,m\omega\sin\omega t+s\cos\omega t\big)\cdot\rho(r,s)drds. 
\end{equation*} 
In particular, 
\begin{itemize} 
	\item 
		$\overline q(t)=\overline r\cos\omega t
		+\overline s\frac1{m\omega}\sin\omega t$, \quad 
		$\overline p(t)=-\overline r\,m\omega\sin\omega t
		+\overline s\cos\omega t$; 
	\item 
		$\overline{\Delta q^2}(t)=\overline{\Delta r^2}
		\cos^2\omega t+\overline{\Delta r\cdot\Delta s}\,
		\frac1{m\omega}\sin2\omega t+{\overline{\Delta s^2}}\,
		\frac1{(m\omega)^2}\sin^2\omega t$; 
	\item 
		$\overline{\Delta p^2}(t)=\overline{\Delta r^2}\,
		(m\omega)^2\sin^2\omega t-\overline{\Delta r\cdot\Delta s}\,
		m\omega\sin2\omega t+{\overline{\Delta s^2}}\cos^2\omega t$. 
\end{itemize} 
Thus, here the conditions of Proposition \ref{SC} are fulfilled with 
\begin{equation*}
	x\!=\!\begin{pmatrix} q \\ p\end{pmatrix}\!,  
	s\!=\!\begin{pmatrix} r \\ s\end{pmatrix}\!, 
	A\!=\!\begin{pmatrix}\cos\omega t\!\!&-\frac1{m\omega}\sin\omega t 
		\\ m\omega\sin\omega t\!\!&\cos\omega t\end{pmatrix}\!,  
	B\!=\!\begin{pmatrix}\cos\omega t\!\!&\frac1{m\omega}\sin\omega t
		\\ -m\omega\sin\omega t\!\!&\cos\omega t \end{pmatrix}\!.
\end{equation*}	
So, here 
$\overline{\Delta x^2}=\overline{\Delta q^2}+\overline{\Delta p^2}
=g_{\mu\nu}\cdot\sigma^{\mu\nu}=\mathrm{Tr}(g(t)\sigma)$, 
where 
\begin{align*} 
	g(t)&=\begin{pmatrix}
	      \cos^2\omega t+(m\omega)^2\sin^2\omega t
	      &\big(\frac1{m\omega}-m\omega\big)\cos\omega t\sin\omega t \\
	      \big(\frac1{m\omega}-m\omega\big)\cos\omega t\sin\omega t
	      &\cos^2\omega t+\big(\frac1{m\omega} \big)^2\sin^2\omega t	
	      \end{pmatrix}, \quad  \det g(t)=1,\\ 
  \sigma&=\begin{pmatrix}
	      \overline{\Delta r^2}&\overline{\Delta r\cdot\Delta s} \\
	      \overline{\Delta r\cdot\Delta s}&\overline{\Delta s^2}
	      \end{pmatrix}, \quad \det\sigma>0. 
\end{align*}
Note, 
$\overline{\Delta x^2}(t)=\overline{\Delta r^2}+\overline{\Delta s^2}$
if $mk=1$. 
\end{exa} 

\begin{exa}\label{ELE} 
In the Lagrangian situation (see Section \ref{LD}), let 
$n=2$, $q=(q^1,q^2)\in\mathbb R^2$, $L(q,\dot q)=(\dot q^1)^2\big/2-U(q)$.  
Then, the Euler-Lagrange equations take form 
$\ddot q^1=-\partial_{q^1}U$, $0=-\partial_{q^2}U$. 
They have the constraint (the solvability condition) $U=U(q^1)$, 
so here the vector field $v=(v^1,v^2)$, where $v^1(q^1)=-\partial_{q^1}U(q^1)$, 
while $v^2(q^1,q^2)$ is arbitrary. In particular, 
we have no reason to expect that the vector field $v$ is divergence free. 
\end{exa} 

\begin{exa} 
Let $x=(x^\mu)\in\mathbb R^m$, $v : \mathbb R^m\to\mathbb R^m$, $x\mapsto x$, 
(in other words, $v=(v^\mu(x))$, $v^\mu(x)=x^\mu$), the divergence
$\mathrm{div\,}v=\partial_{x^\mu}x^{\mu}=m$, 
the total time derivative $D_t=\partial_t+x^\mu\partial_{x^\mu}$. 
In this case, the Liouville equation 
\begin{equation*}
	\partial_t R+\partial_{x^\mu}\big(x^\mu\cdot R\big)+\mathrm{div\,}v\cdot R
		=D_t R+m\cdot R=0,
\end{equation*} 
has the general solution 
\begin{equation*} 
	R(t,x)=R(t,x^1,\dots,x^m)
	=\mathrm{e}^{-mt}\rho(s)=\mathrm{e}^{-mt}\rho(s^1,\dots,s^m), 
\end{equation*}
where $\rho(s)$ is an arbitrary function, we assume $\rho\ge0$ 
and $\int\rho(s)d^ms=1$, while 
\begin{equation*} 
	s=x\mathrm{e}^{-t}, \quad (x=s\mathrm{e}^t) 
\end{equation*} 
are the solutions of the equation $D_t s=0$. 
The average value of a function $f(t,x)$ here is written in the form 
\begin{equation*} 
	\overline f(t)=\int f\big(t,s\mathrm{e}^t\big)\rho(s)d^ms. 
\end{equation*} 
In particular, 
\begin{align*} 
	&\overline x(t)=\int s\mathrm{e}^t\rho(s)d^ms
	=\mathrm{e}^t\cdot\overline s, \\
	&\overline{\Delta x^2}(t)
	=\int\big(s\mathrm{e}^t-\overline s\mathrm{e}^t\big)^2\rho(s)d^ms
	=\mathrm{e}^{2t}\cdot\overline{\Delta s^2}. 
\end{align*}
\end{exa} 

\section{Conclusion.} 
From the above considerations and simple examples one can see 
that functional mechanics has the solid mathematical foundations. 
Thus, the problem left is the metaphysical foundations of its relevance 
to the physical laws and the experimental interpretation. 
In particular, one should have the clear answers to the 
natural questions listed below. 
\begin{itemize} 
	\item 
		What is the origin of the vector field $v : \mathbb R^m\to\mathbb R^m$ 
		in the conception of the functional mechanics? 
		Is it governed by the standard classical mechanics, 
		Hamiltonian or Lagrangian? 
	\item 
		If the vector field $v$ is governed by the Hamiltonian 
		mechanics, then it is divergence free, i.e., $\mathrm{div\,}v=0$, 
		(see Proposition \ref{DF}), the wave packet $R(t,x)=\rho(S(t,x))$,
		where the profile $\rho(s)$ is arbitrary while the flow $s=S(t,x)$ 
		is given by the first integrals of the Liouville equation. 
		One may interpret this situation as follows: 
		the dynamics is governed by the usual Hamiltonian equations, while 
		the profile $\rho$ characterises measurement.  
	\item 
		If the vector field $v$ is governed by the Lagrangian mechanics, 
		then there is no reason to expect that the vector field 
		$v$ is divergence-free (see Example \ref{ELE}), so the dynamics 
		is governed directly by the Liouville equation (\ref{LE}), 
		the wave packet $R=R(t,x)$ does not split into the profile and 
		the flow. 
	\item 
		Most mathematical models of the modern mathematical physics 
		are subject to the Lagrangian approach and are formulated as  
		extremals of the appropriate functionals (actions). 
		It is naturally requires the applicability of the functional 
		mechanics to variational problems. 
\end{itemize} 
Clear, this list is far from to be complete, so the functional mechanics 
has a long way ahead. Let us hope it will be fruitful and successful.


\begin{thebibliography}{99}

\bibitem{IVV1}
	Igor V. Volovich, 
	Time Irreversibility Problem and Functional Formulation of 
	Classical Mechanics, arXiv:0907.2445v1. 
	
\bibitem{TV} 
	A. S. Trushechkin and I. V. Volovich, 
	Functional Classical Mechanics and Rational Numbers, 
	P-Adic Numbers, Ultrametric Analysis, and Applications, 
	1:4 (2009), 361--367, arXiv:0910.1502v1. 
	
\bibitem{IVV2} 
	Igor V. Volovich,
	Randomness in Classical Mechanics and Quantum Mechanics, 
	Found. Phys., 41:3 (2011), 516--528, arXiv:0910.5391v1. 
	
\bibitem{IVV3} 
	I. V. Volovich, 
	Functional mechanics and time irreversibility problem, 
	Quantum bio-informatics III, QP–PQ: Quantum Probab. 
	White Noise Anal., 26, World Sci. Publ., Hackensack, NJ, 
	2010, 393--404. 
	
\bibitem{IVV4}
	 I. V. Volovich, 
	 Bogoliubov equations and functional mechanics, 
	 Theoret. and Math. Phys., 164:3 (2010), 1128--1135. 
	 
\bibitem{IVV5} 
	I. V. Volovich, 
	Functional stochastic classical mechanics, 
	P-Adic Numbers Ultrametric Anal. Appl., 7:1 (2015), 56--70. 
	
\bibitem{Z} 
	V. V. Zharinov, 
	Binary relations, B\"acklund transformations, and wave packet 
	propagation, 
	Theoret. and Math. Phys., 205:1 (2020), 1245--1264.
\end{thebibliography}
\end{document}